\documentclass[aps,prb,twocolumn,showpacs,superscriptaddress,groupedaddress]{revtex4}
\usepackage{epsfig,amsopn}
\usepackage{graphicx}
\usepackage[usenames]{color}
\usepackage{amsmath,amssymb}
\usepackage{enumerate}
\newcommand\bea{\begin{eqnarray}}
\newcommand\eea{\end{eqnarray}}
\newcommand\beq{\begin{equation}}
\newcommand\eeq{\end{equation}}
\newcommand{\noi}{\noindent}
\newcommand{\bib}{\bibitem}
\def\nn{\nonumber}
\def\f{\frac}

\def\ep{\epsilon}

\def\si{\sigma}
\def\Do{\partial}
\def\De{\Delta}
\def\dg{\dagger}

\def\ua{\uparrow}
\def\da{\downarrow}

\def\inf{\infty}

\begin{document}
\title{Enhancement of crossed Andreev reflection in a superconducting ladder
connected to normal metal leads}
\author{Abhiram Soori}
\affiliation{International Centre for Theoretical Sciences, 
Tata Institute of Fundamental Research, Survey~No.~151, 
Shivakote, Hesaraghatta Hobli, 
Bengaluru North 560089, India}
\author{Subroto Mukerjee}
\affiliation{ Department of Physics,
Indian Institute of Science, Bengaluru 560012, India }
\begin{abstract} 
Crossed Andreev reflection~(cAR) is a scattering process that happens in
a quantum transport set-up consisting of two normal metals~(NM) attached to
a superconductor~(SC), where an electron incident from one NM results 
in a hole emerging in the other. Typically, electron tunnelling~(ET) 
through the superconductor from one NM to the other competes with cAR and
masks its signature in the conductance spectrum. We propose a novel
scheme to enhance cAR, in which the SC part of the NM-SC-NM  is side-coupled to
another SC having a different superconducting phase to form a Josephson junction in the
transverse direction. At strong enough coupling and for a large enough phase difference,
one can smoothly traverse between the highly ET-dominant to
the highly cAR-dominant transport regimes by tuning chemical potential, 
due to the appearance of subgap Andreev states that are extended in 
the longitudinal direction. We discuss connections to realistic systems.
\end{abstract}

\maketitle
\section{Introduction}
Andreev reflection~(AR) is the scattering process by which a current 
flows across the interface of a normal metal~(NM) and a 
superconductor~(SC) when an external bias across the interface is
applied in the subgap regime. The Cooper-pair current in the superconductor
draws equal contributions from the electron and hole channels of the normal
metal. This phenomenon, first discovered by Andreev~\cite{andr64} has been
extensively studied theoretically and experimentally for several
decades~\cite{btk,kasta,mfexp}. 

AR has played an important role in the observation of transport signatures of exotic Majorana fermions in mesoscopic
systems~\cite{mfexp}. 
In addition, recent advances in 
cold-atomic systems promise new testbeds where theoretical
findings~\cite{ar-like,ochamon06} related to Andreev reflection can
be demonstrated experimentally~\cite{domanski11}.
Crossed Andreev reflection~(cAR) is a phenomenon closely related to AR
and occurs in a system consisting of two normal metals attached to a 
superconductor~\cite{melin09,beckmann04,kueisun,melin04,yamas03a,rein13,
yeyati-em,he13,car-graphene,
chen15,cht03,russo05,byer95,deuts00}. 
An electron incident on the SC from the first~normal~metal~(NM1) gets
absorbed into the SC as a Cooper-pair, absorbing the second electron 
of the Cooper-pair from the second~normal~metal~(NM2), resulting in a hole 
current in NM2.
The phenomenon of cAR is closely related to the production of non-locally 
entangled electrons by splitting Cooper pairs from the SC, the detection 
and enhancement of which has seen a lot of theoretical~\cite{cps-th} and 
experimental interest~\cite{cps-das,cps-schindele}.
However, cAR is accompanied by electron tunneling~(ET) where the electron 
from NM1 tunnels into NM2 as an electron. cAR  is typically masked
in simple NM-SC-NM systems due to dominant ET~\cite{melin09}. A negative
differential transconductance between NM1 and NM2 is a definite signature 
of cAR. 

In this paper, we propose a novel scheme to enhance cAR, which is different 
from other existing proposals~\cite{melin04,yamas03a,deuts00,rein13,
chen15,car-graphene,he13,yeyati-em,
cht03}, and demonstrate with a simple theoretical model that 
cAR enhancement can be much greater than predicted in them. Of the two existing proposals that
have been experimentally realized, the first method introduces barriers at the NM-SC junctions of an NM-SC-NM 
set-up~\cite{cht03,russo05}, 
while the second employs two ferromagnets~(FM) in an anti-parallel configuration
instead of the NM's~~\cite{beckmann04,melin04,yamas03a} that suppress ET and AR 
thereby allowing cAR to dominate. 
cAR is enhanced in the former method when the momentum scale characterizing the
barrier is at least as large as the Fermi-momentum, though the enhanced cAR
currents are too small as shown in appendix A. In addition, there are several other proposals
such as: 
(i)~employing quantum spin Hall insulators connected to SC and spatially separate
ET and cAR channels based on the spin-momentum locking of the edge states~\cite{rein13}, 
(ii)~driving a steady Cooper pair current in the SC such that the SC~phase
modulates and thus enhances cAR~\cite{chen15}, 
(iii)~substituting NM and SC parts of the setup with exotic materials such as 
graphene, silicene, topological insulators and topological superconductors~\cite{car-graphene,he13}, 
(iv)~coupling to external electromagnetic modes~\cite{yeyati-em}, 
which have not yet been realized experimentally.

\begin{figure}[htb]
\includegraphics[width=8.6cm]{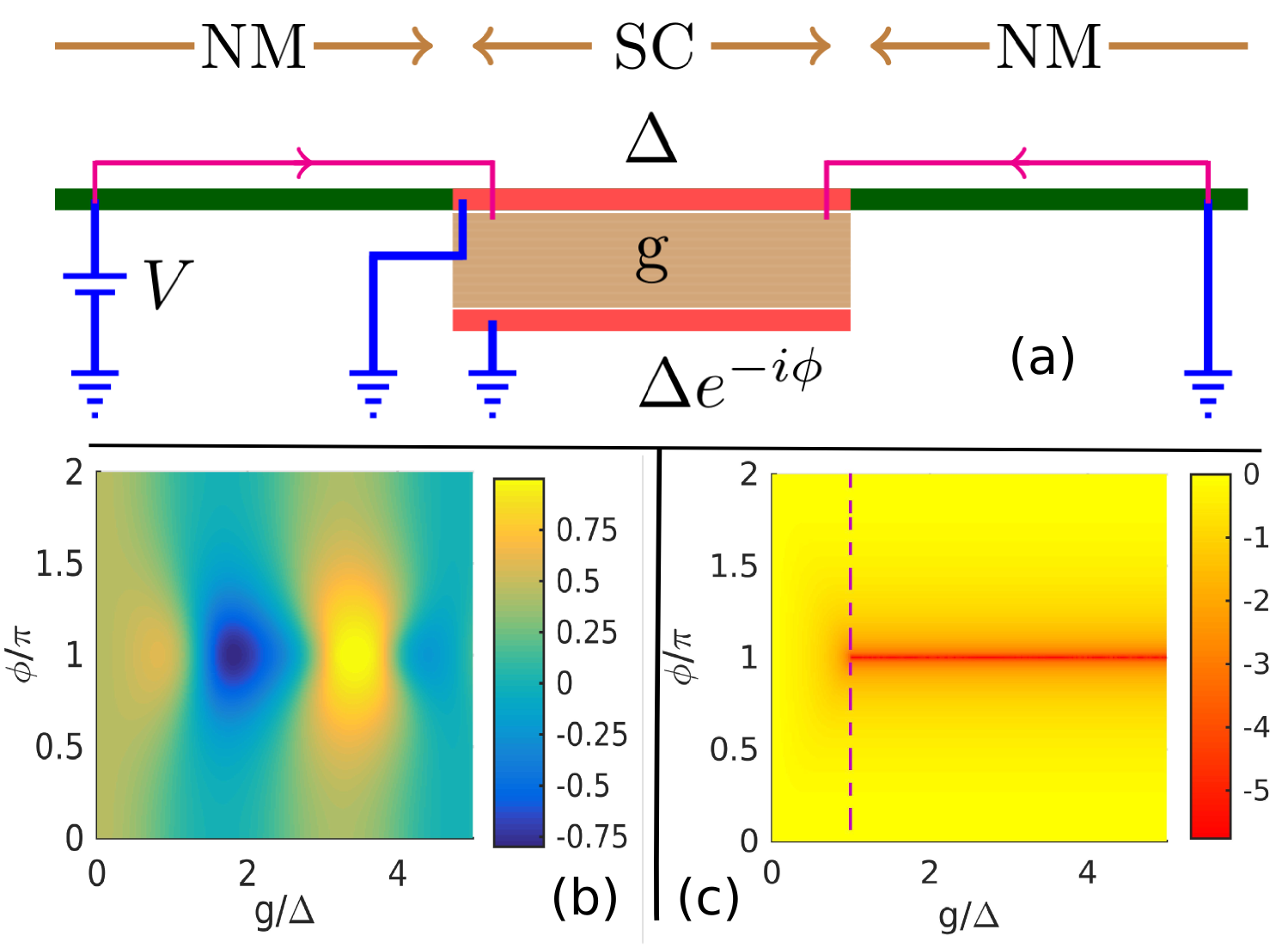}
\caption{ (a) A schematic diagram of the set-up. The thick red lines in the middle 
are the two SC-channels, coupled to each other by a strength $g$-highlighted by the 
brown region in between. A voltage bias $V$ is applied to the left NM while 
the SC-ladder and the right NM are grounded, as shown by the blue lines. Thin 
lines with arrows show the current directions when cAR dominates over ET.
(b) The zero bias transconductance $G_{RL}$ in units of $2e^2/h$.
(c) The logarithm of the gap in the dispersion of the ladder region 
[Eq.~\eqref{eq-disp}] in units of $\Delta$; the vertical dashed line corresponds to 
$g/\De=1$. The parameters chosen are $\mu=10\Delta$, $a=6\hbar/\sqrt{2m\De}$.}
\label{fig-cont-phi-g}
\end{figure}

Here, we propose to modify the NM-SC-NM setup by side-coupling the SC 
with another SC (which has a supercondcuting phase differing by $\phi$ but the same 
magnitude of the pair potential) as shown in Fig.~\ref{fig-cont-phi-g}(a).
We call the two coupled SC's together an `SC~ladder' {\sl (note that this is 
different from several setups such as the one by Grosselin~et~al~\cite{gosselin14}
where a magnetic flux enclosed between two superconductors forming a loop can control 
subgap transport in the superconductors)}. We show that an
adequate supercondcuting phase difference between two legs of the ladder 
accompanied by a sufficiently strong coupling between two legs of the
SC~ladder leads to subgap Andreev states which can enhance cAR. 
This is the central result of our work. 

 \section{Details of the calculation}~\label{sec-calc}
 \subsection{Hamiltonian and dispersion}
Introducing the hole annihilation operators  $d_{s,\lambda}(x) = 
c^{\dg}_{\bar{s},\lambda}(x)$, the Hamiltonian  for the ladder 
region ($0\le x\le a$) can be  written as:
\bea  H &=& \sum_{\lambda,s}\Bigg[\Psi^{\dg}_{s,\lambda}(x) 
\Big\{ \Big(\f{-\hbar^2}{2m} \f{\Do^2}{\Do x^2} -\mu\Big)\tau_z +
\Delta (\cos{\phi_{\lambda}}\tau_x \nn \\ && ~~+ \sin{\phi_{\lambda}}
\tau_y) \Big\}\Psi_{s,\lambda}(x) + g~\Psi^{\dg}_{s,\lambda}(x) 
\tau_z\Psi_{s,\bar{\lambda}}(x)\Bigg],  ~\label{ham02} \eea
where (i) $s = \uparrow,\downarrow$ and $\lambda=1,2$ 
~are~spin and wire~indices respectively, (ii)~$\bar{s} / \bar{\lambda}$ 
takes the~value of~the label~index different from $s/\lambda$, and 
(iii)~$\Psi_{s,\lambda}(x)=[\si_s c_{s,\lambda}(x),~d_{s,\lambda}(x)]^T$, 
where~$\si_{\ua/\da}=\pm 1$.
The dispersion for this Hamiltonian is: 
\beq E = \pm\sqrt{\epsilon_k^2+g^2+\De^2\pm2g\sqrt{\epsilon_k^2
+\Delta^2\sin^2{[(\phi_1-\phi_2)/2}]}}, \label{eq-disp} \eeq 
where $\ep_k =(\hbar^2k^2/2m-\mu)$ and $\hbar k$ is the momentum. Each
of these bands is doubly degenerate due to spin ($s=\uparrow,\downarrow$).
The Hamiltonians for the NM~regions on the left
($x<0$) and right~($x>a$) are $H=\sum_s c^{\dg}_{s}(x)
[\hbar^2 k^2/2m-\mu]c_{s}(x)$. From here onwards, we shall leave out the spin
index~$s$ and a factor of $2$ to account for spin degeneracy will multiply 
the conductances in the final result. 
 The superconducting phases $\phi_{\lambda}$ on the two legs ($\lambda=1,2$) of the
 SC~ladder can be chosen to be $\phi_1=0$, and $\phi_2=-\phi$, 
 without loss of generality so that 
$(\phi_1-\phi_2)=\phi$. 

 \subsection{Wavefunction}
The solution to the Hamiltonian 
is a two-spinor $\psi$ in the NM-regions (the two components represent 
the electron- and hole- channels) and a four-spinor $[\psi^T, ~\chi^T]^T$
in the ladder region (two more components represent the second leg of 
the ladder),  
where the spinors $\psi$ and $\chi$ have the form: 
$\psi(x)=[\psi_e(x),~\psi_h(x)]^T$ for $-\infty<x<\infty$ and
$\chi(x)=[\chi_e(x),~\chi_h(x)]^T$ for $0\le x\le a$. 
In the NM regions, the electrons and holes have momenta $\pm \hbar k_e$ and 
$\pm\hbar k_h$ respectively, where $k_{e/h}=\sqrt{2m(\mu\pm E)/\hbar^2}$ 
at a given energy $E$. The scattering wavefunction  for an electron incident
from the left onto the ladder region at an energy $E$ is given by: 
\bea \psi_e(x) &=& e^{i k_e x}+r_n e^{-i k_e x}, {\rm~~for~~}x\le 0, \nn \\
 &=& t_n e^{i k_e x}, {\rm~~for~~}x \ge a, \nn \\
 \psi_h(x) &=& r_a e^{i k_h x}, {\rm~~for~~}x\le 0, \nn \\
  &=& t_a e^{-i k_h x}, {\rm~~for~~}x \ge a, \nn \\
  \psi(x) &=& \sum_{\si,\nu,p} s_{\si,\nu,p} e^{i\si k_{\nu,p} x} 
[\psi_{e,\nu,p}~,~\psi_{h,\nu,p}]^T~ {\rm and} \nn \\
\chi(x) &=& \sum_{\si,\nu,p} s_{\si,\nu,p} e^{i\si k_{\nu,p} x} 
[\chi_{e,\nu,p}~,~\chi_{h,\nu,p}]^T, \nn \\ 
&& {\rm for}~~0\le x\le a.\label{eq-wf1}
\eea

In the ladder region, the momenta $\si \hbar k_{\nu,p}$ at energy
$E$ denoted by the indices $\si,\nu,p$  are obtained by inverting
the dispersion relation Eq.\eqref{eq-disp}:
$\hbar k_{\nu,p} =\sqrt{2m(\si_p \ep_{k,\nu}+\mu)}$, where
$\ep_{k,\nu} = \sqrt{E^2+g^2 -\De^2 + 
2g \nu \sqrt{E^2-\De^2\cos^2(\phi/2)}}$,
and the index $\si=\pm 1$ denotes whether the mode is for a right/left-mover, the 
index $\nu=\pm 1$  refers to the anti-bonding/bonding bands formed due to the
hybridization between the two legs ($\lambda=1,2$) of the ladder 
and $p~=~e,h$ refers to electron,~hole~-like bands for which
$\si_{e/h}~=~\pm1$. 

\subsection{Boundary conditions}
From the boundary conditions~(BC's) described by Carreau et.~al~\cite{carreau90}
for a general one dimensional system, we choose the one that is physically 
relevant to our system.
The NM-SC junction generically has a barrier for
electron tunneling which limits the electron transmission. The BC at the NM-SC
interface is described by the continuity of the wave-function and a
discontinuity of the derivative~\cite{btk}. The latter is the same as 
having a delta-function barrier potential on the NM side of the NM-SC
junction, infinitesimally close to the junction. 
Since we are interested 
in enhancing cAR over ET, we set the barrier strengths to zero to make
the NM-SC junctions fully transparent. In other words, both the wavefunction
and its derivative at the NM-SC interface are continuous. But this fixes
the BC only for one leg of the ladder. The BC for the other leg is given by
a probability current conserving BC at the ends of the ladder (i.e., 
$x=0,~a$). Such a BC that describes the lower leg of the ladder depends on 
four-parameters in general as for the `particle~in~box' study of Carreau
et.~al~\cite{carreau90}. The NM-SC interfaces at $x=0,~a$ are not connected
by any direct hopping as in the case of periodic BC's. This causes the BC's 
to depend on just two parameters:
$(q_{2,x_0}+\Do_x)\Psi_{2}(x)|_{x=x_0}=0$, where $q_{2,x_0}$~is a real-valued
parameter with dimensions of inverse length that describes the BC at $x_0$. 
The limit $q_{x0}\to \infty$ implies that the wavefunction is 
zero at $x=x_0$, while the limit $q_{x0}\to 0$ implies that the first 
derivative of the wavefunction is zero at $x_0$, allowing the wavefunction
to have a non-zero probability density at $x_0$.  The latter limit 
qualitatively corresponds to a lattice model of the SC ladder in which the last
site of the second leg can have a finite probability density.  Hence, we 
have choose $q_{x_0}=0$ for both $x_0=0$ and $x_0=a$, though a particular
choice of the BC does not affect the results qualitatively. The 
BC's used are: 
\bea  \psi(x_0^+)=\psi(x_0^-)~, && \Do_x\psi(x)|_{x=x_0^+}=\Do_x 
\psi(x)|_{x=x_0^-}~, \nn \\ \Do_x \chi(x)|_{x=x_0} = 0 && 
{\rm ~for~} x_0=0,~a. \label{eq-bc-s}  \eea

Both the NMs are connected to only one leg~($\lambda=1$) of the SC~ladder 
as shown in Fig.~\ref{fig-cont-phi-g}(a). As mentioned earlier, the junction is modeled to be 
transparent since we are interested in enhancing cAR - a scattering process
which involves both the NMs. We first calculate the scattering amplitudes
numerically for a diverse set of relevant parameters by employing the 
BC-~eq.\eqref{eq-bc-s} in the wavefunction given by Eq.\eqref{eq-wf1}. Using
these scattering amplitudes, the transconductance is calculated, 
which is the physical quantity of interest since it can be experimentally measured. 

\subsection{Transconductance}
The system under investigation is essentially a three terminal set-up. 
The NM regions on the left and right (extending to $ x \to \mp \infty$)
form two terminals and the SC ladder in the middle (maintained at a fixed 
chemical potential $\mu$) acts as a reservoir for the charge current 
since it is not conserved in the SC region. 

We are interested in calculating the differential transconductance
$G_{RL}(V) := ~dI_R(V)/dV$ - where $dI_R(V)$ is 
the change in current in the right NM when the Fermi energy of the left 
NM is changed from $eV$ to $e(V+dV)$, keeping the Fermi energies of the 
ladder region and right NM at zero (here, $e$~is the electronic charge 
and $V$ is the voltage). Using the Landauer-B\"uttiker 
formalism~\cite{landau-butti}, the differential transconductance in 
such a transport set-up
at a bias voltage~$V$ is  
\beq G_{RL} (V) =  \f{2 e^2}{h} \Big(|t_n|^2 -  |t_a|^2 
\sqrt{(\mu - eV)/(\mu+eV)} ~\Big) \label{eq-grl}\eeq
It is easy to see from Eq.\eqref{eq-grl} that the contributions of  
ET and cAR to the current on the right NM are positive and negative 
respectively. Hence, a negative  $G_{RL}$ is a clear signature of 
 cAR enhanced over ET while a positive $G_{RL}$ implies the dominance 
of ET over cAR. 

\section{ Results} 
\begin{figure}[htb]
\includegraphics[width=8.4cm]{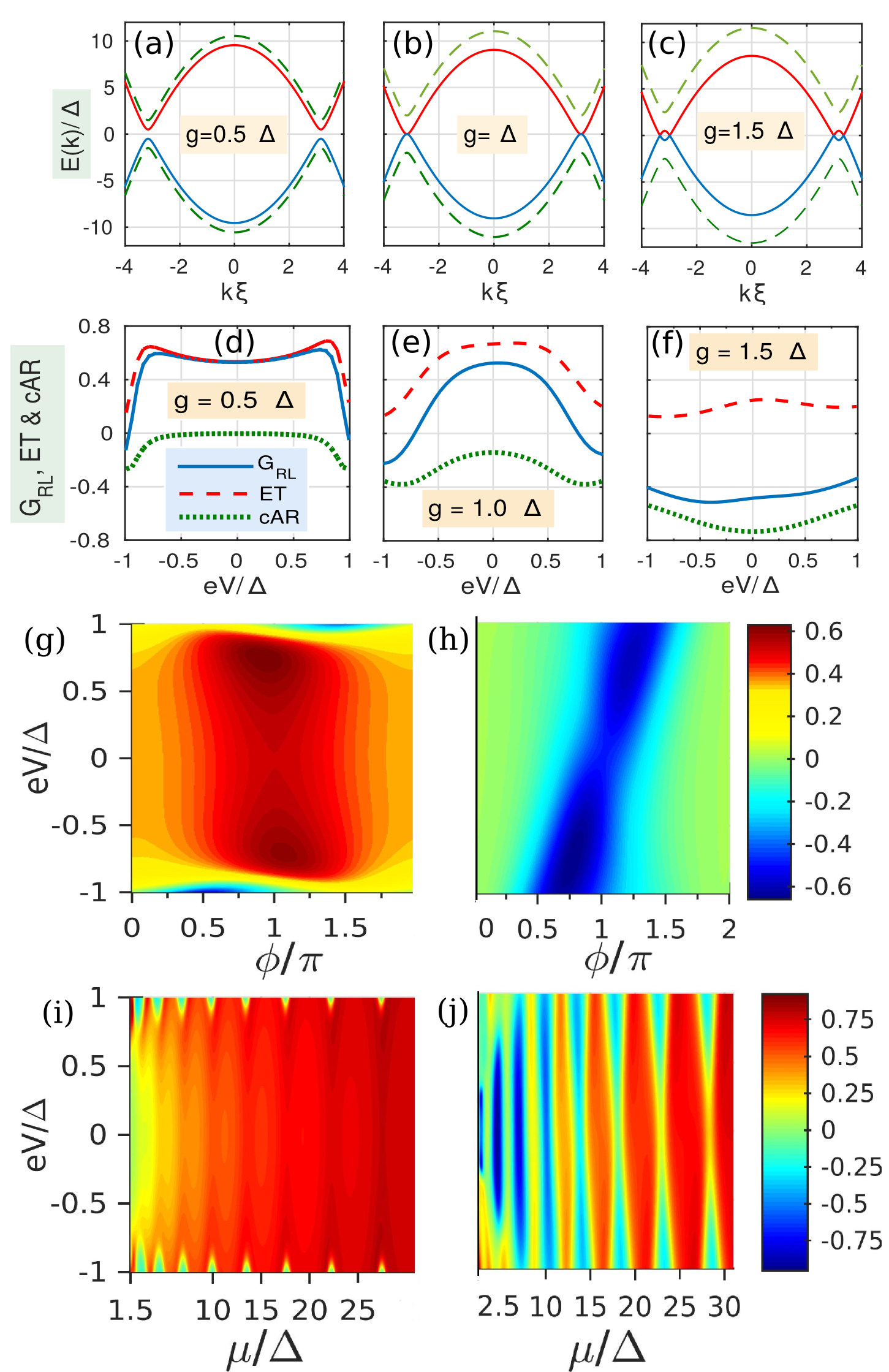}  
\caption{ (a-c):~Dispersion of the SC ladder,  
(d-f):~transconductance $G_{RL}$, contributions to it from ET and cAR
for $\phi=\pi,~\mu=10\De$.
  (g-j):~Contour plots of $G_{RL}$. 
  For~(g,~i), $g=0.5\De$. For~(h,~j), $g=1.5\De$. 
  For (d-j), $a=6\hbar/\sqrt{2m\De}$. For~(g,~h), $\mu=10\De$.
  For~(i,~j),~$\phi=\pi$. $G_{RL}$ is units
  of $2e^2/h$.}
  \label{fig-E-phi-mu}
\end{figure}

(1)~The zero-bias transconductance
$G_{RL}|_{V=0}$ as a function of the parameters $\phi$ and $g$ is
shown in  Fig.~\ref{fig-cont-phi-g}~(b) 
and understand it in terms of the dispersion of the ladder region given by 
Eq.\eqref{eq-disp}. Fig.~\ref{fig-cont-phi-g}~(c) is a contour-plot of the
logarithm of the gap in units of $\De$ (i.e.,~$\log{[E_g/\De]}$). The gap
closes on the line: $g/\De \ge 1, ~\phi=\pi$. cAR is enhanced as 
$g/\De$ crosses $1$ from left to right and this enhancement is prominent
around $0.7\pi<\phi<1.3\pi$ and $1.4\De < g < 2.2\De$. Further, as $g/\De$
crosses a value of $3$, ET is enhanced to a value as high as 
$\sim 0.75\cdot 2e^2/h$. This indicates that the enhancement of
cAR and ET to such high values is related to closing of the gap. 
\\ (2)~Fixing $\phi=\pi$, we see how the subgap conductance spectrum 
changes as the coupling strength~$g$ is changed.  This is contrasted
with the dispersion of the ladder region, the topology of which changes
when $g/\De$ is on either side of~$1$. In Fig.~\ref{fig-E-phi-mu}~(a-c),
the dispersion of the SC ladder, and in Fig.~\ref{fig-E-phi-mu}~(d-f), 
the conductance spectrum for $g/\De=0.5,~1.0,~1.5$ have been plotted.
As $g/\De$ crosses over from $0.5$ to $1.5$, through $1.0$, the 
conductance spectrum shifts smoothly from being in the positive-half
to being in the negative-half of the ordinate, which is a clear 
signature of enhanced cAR as indicated by the green dotted lines. 
This enhancement is accompanied by the appearance of subgap Andreev states in 
the SC~ladder.
\\ (3)~The dependence of the conductance spectrum in the bias
window ($|eV|\le\De$) on the superconducting phase difference $\phi$ and the chemical potential $\mu$ can be obtained. 
Fig.~\ref{fig-E-phi-mu}~(g-h) shows contour plots of $G_{RL}$ as function
of $(eV, \phi)$ 
for $g=0.5\De$ [\ref{fig-E-phi-mu}(g)] and $g=1.5\De$ [\ref{fig-E-phi-mu}(h)].
One can see that 
cAR is enhanced significantly for the case $g=1.5\De$ throughout in the
bias window $(-\De,\De)$, and the value of $\phi$ at which cAR is 
enhanced the most depends on the value of the bias $eV$. Though cAR is enhanced
in the case of $g=0.5\De$, this happens in very narrow regions of the 
$eV-\phi$ plane and it is worth noting that the enhancement of cAR for
$g=0.5\De$  happens for values of $\phi$ far away from zero or $2\pi$. \\
(4) Fig.~\ref{fig-E-phi-mu}~(i-j) show contour~plots of $G_{RL}$ 
as a function of $(eV,\mu)$ for $g=0.5\De$ [\ref{fig-E-phi-mu}(i)]
and $g=1.5\De$ [\ref{fig-E-phi-mu}(j)]. For these two cases, 
$\phi=\pi$. It is apparent from the contour-plots that the enhancement
of cAR is prominent in the case of $g=1.5\De$ compared to the case 
of $g=0.5\De$. For $g=1.5\De$, cAR is enhanced throughout the bias 
window around certain values of $\mu$, but for $g=0.5\De$, a poor cAR 
enhancement can be found in very small regions near $eV=\pm\De$. 
Further, the enhancement of cAR occurs at a series of $\mu_i$'s. 
Also, the enhancement of cAR around the $\mu_i$'s becomes less prominent with
increasing $\mu_i$.

\section{ Mechanism } 
The above results point to the following mechanism for
the enhancement of cAR: A nonzero phase-difference $\phi$ and a non-zero
coupling $g$ between the two legs of the ladder create plane-wave modes 
within the otherwise gapped dispersion of the ladder region as can 
be seen from Eq.\eqref{eq-disp}. Equivalently, two out of four 
$k_{\nu,p}$'s become real valued at energies: $|E|<\De$. These 
plane wave modes have nonzero components in both
  electron- and hole- sectors and we call
them subgap Andreev states. Thus, the transconductance is due to 
a Fabry-P\'erot type interference between the subgap Andreev states in
 the ladder region that transmit either an electron or 
a hole into the right NM. The probabilities of the two transmission processes can be tuned by 
changing a parameter that changes the dispersion and the spinor 
structure of the modes in the ladder region. The subgap Andreev
states can be thought of as plane wave modes formed by the hybridization
of Andreev bound states, when a large number of Josephson 
junctions (each formed between two superconducting quantum dots) are
coupled.

Since, the chemical potential $\mu$ sets the fundamental length-scale of 
the problem and affects the spinor structure of the BdG modes, changing
$\mu$ smoothly,  keeping the  values of $eV$, $\phi$~($\neq 0$) and
$g$~($\neq 0$) fixed must show a smooth transition from maximally 
enhanced cAR to maximally enhanced ET.  This can be seen in 
Fig.~\ref{fig-E-phi-mu}(j) and the recurrent enhancement of cAR and ET 
at a series of values of $\mu$ ($=\mu_i$) corresponds to a periodicity
in $k_i$'s (where $k_i=\sqrt{2m \mu_i/\hbar^2}$)~~\cite{expl06}, given 
by $(k_{i+1}-k_i)a \sim \pi$, 
reaffirming quantitatively our explanation that the cAR 
enhancement is due to Fabry-P\'erot interference~\cite{fabpero} between the subgap
Andreev states in the SC~ladder. As can be seen from
 Fig.~\ref{fig-cont-phi-g}(c), 
$\phi=\pi$ and $g>\De$ give rise to gap closing and hence the 
cAR enhancement is expected to be maximal for this choice of parameters, 
which is in agreement with the results highlighted in 
Fig.~\ref{fig-E-phi-mu}~(d-j).

\section{Discussion}
We have primarily studied a one-dimensional superconductor
side-coupled to another one dimensional superconductor having a superconducting 
phase different from the first superconductor. However,
long-range superconducting order is not possible in purely one dimension.
This limitation can be overcome by a proximate higher dimensional superconductor in
contact with the one-dimensional quantum wire. Also, the Fermi energy and 
chemical potential in the SC are assumed to be maintained at particular
values, which means that a steady state current flows into the SC and the 
SC is grounded via a grounding electrode.

We now discuss different systems that are closer to experimental setups 
 in which our results can possibly be tested. To be able to maintain a SC~phase
 difference $\phi$ between two legs of the ladder is an important 
 task when it comes to experimental implementation. We make the following 
 proposals to achieve this.\\ (1)~Passing a 
 supercurrent in the transverse direction; a SC~phase 
difference between two legs of the SC ladder is induced that is 
proportional to the transverse current~\cite{chen15}. \\
(2)~Using magnetic flux to mimic the 
Josephson phase difference as in SQUIDS~\cite{tink}. Connecting two
legs of the ladder through a loop and passing a magnetic flux through
can induce a phase difference proportional to the flux 
(modulo flux quantum $\Phi_0$) as discussed in ref.~\cite{anna13}.\\ 
(3)~Using $\pi$-junction materials, such as layered superconductors~\cite{liu95}
which have a non-zero superconducting phase difference naturally existing between the
adjacent layers. Sandwiching such a layered SC between NM~leads in such
a way that each layer lies in the longitudinal direction can mimic the
ladder structure proposed here. This will be a quasi-two-dimensional 
version of the setup we have proposed. We have performed transport 
calculations for  such  a two-dimensional version of the ladder 
geometry and we see that most of our results we obtained above 
remain the same qualitatively (see appendix B). \\
(4) Using a closely related system in which the subgap Andreev states that appear in the ladder also appear namely {\sl a Josephson junction between 
two-dimensional superconductors}. To describe the system more 
 precisely, let the regions:
$(0\le y \le \inf,~0<x<a)$  and  $(-\inf<y<0,~0<x<a)$ have SC 
phases $\De$ and $\De e^{-i\phi}$. Using appropriate BC at $y=0$, the
subgap states localised at the junction can be calculated (see appendix B).
We see that very similar to the ladder, a non-zero $\phi$ gives
rise to subgap states which are BdG planewave modes along $x$-direction,
but are localised in the $y$-direction around $y=0$. These states 
can be used to enhance cAR if two NM metal leads are connected 
close to the Josephson junction.

The very fact that a phase difference is maintained between the two
legs of the ladder, means that a Josephson current flows from one 
leg of the ladder to the other in the transverse direction. However, 
this current does not interfere with the quasiparticle current that 
is carried by the subgap Andreev states between the two NM leads. 
In a recent experiment~\cite{pillet10} , it was demonstrated that a subgap Andreev 
bound state formed in a Josephson junction shows a subgap peak in 
conductance when connected to  a NM. This is due to Andreev reflection
and the current in the NM due to Andreev reflection does not interfere 
with the Josephson current that flows from one SC to the other. 
The existence of stabilized subgap modes is required to obtain 
enhanced cAR.

In a set-up where cAR is enhanced over ET and AR, if the bias is maintained
across both the NM-SC junctions keeping the two NM's grounded, this a high-efficiency Cooper pair splitting~(CPS) 
results. Once a considerable superconducting phase difference ($\pi/2 \lesssim \phi \lesssim 3\pi/2$)
is maintained in the ladder, CPS can be  enhanced by tuning either the 
chemical potential $\mu$ or the length $a$ of the ladder region, in 
contrast to the already existing Cooper pair splitter~\cite{cps-das,
cps-schindele} which is based on Coulomb blockaded quantum dots, where two gate
voltages need to tuned to a particular combination. In this respect, 
our scheme may be more robust to the parameters that need to be tuned
to get CPS enhancement. Further, the conductances 
measured in the experiment reporting high efficiency
CPS~\cite{cps-schindele} suggest that the corresponding cAR enhanced 
transconductance values are much smaller in magnitude than the values
that can be obtained theoretically in our set-up.

\begin{acknowledgements}
AS thanks Diptiman Sen, G Baskaran, Anindya Das, Oindrila
Deb, Aviad Frydman, Efrat Shimshoni, Pascal Simon, Ady Stern, Manisha
Thakurati,  Aveek Bid, Hemanta Kundu  and Alfredo Levy Yeyati 
for discussions. AS thanks Abhishek Dhar for
support and encouragement and {\it Infosys Excellence Grant
for ICTS-TIFR, Bengaluru} for funding expenses to participate in 
{\sl APS March meeting 2016}, where this work was presented.
We thank  Diptiman~Sen for careful reading of the manuscript.
\end{acknowledgements}

\appendix

\section{NM-SC-NM with barriers} 
\begin{figure}[htb]
  \includegraphics[width=8cm]{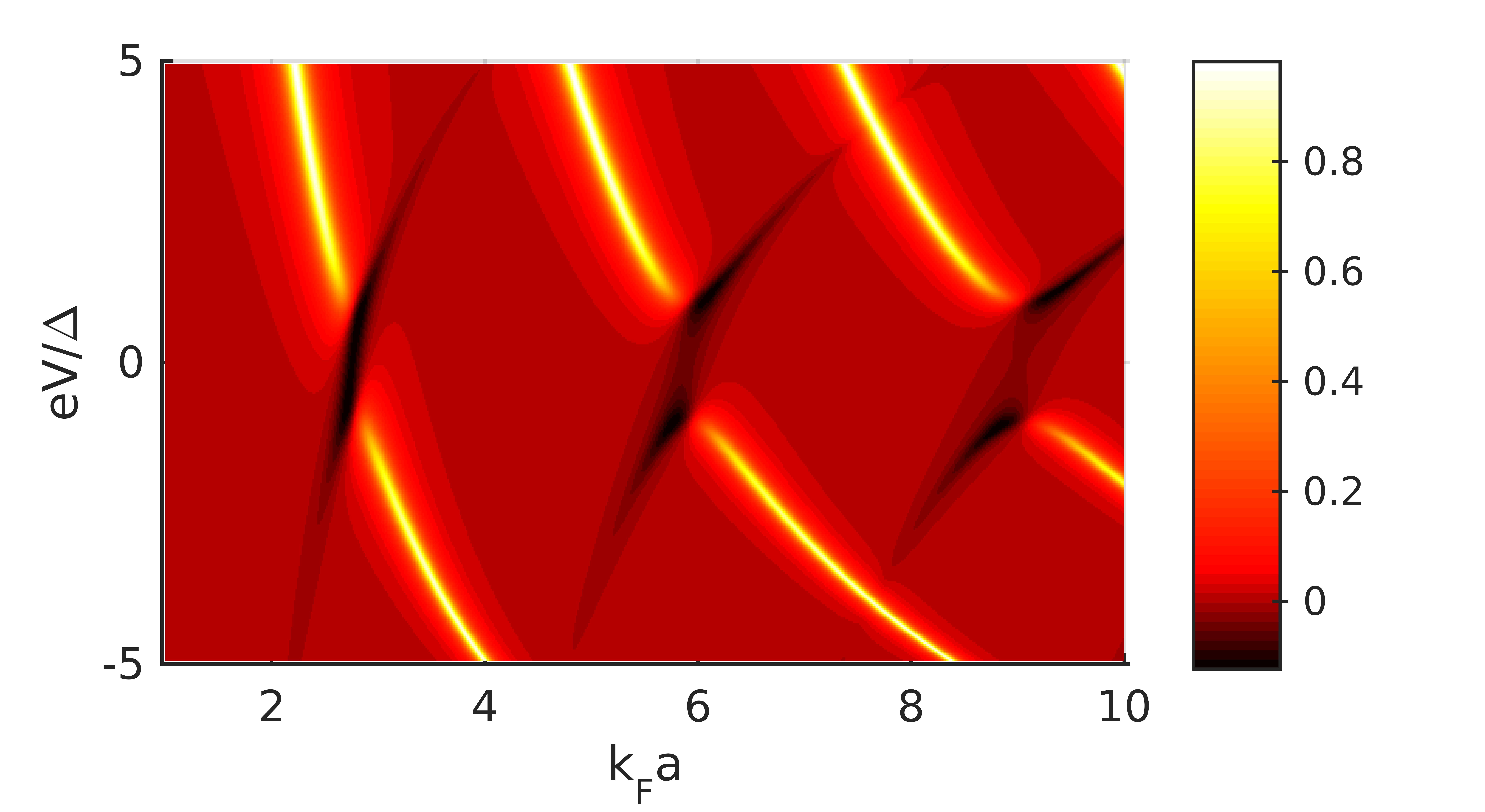} 
  \includegraphics[width=8cm]{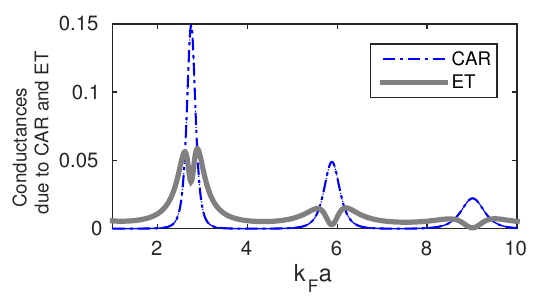}
  \caption{Our calculation of ET and cAR contributions in a single channel
  NSN geometry with delta-function barriers of $q_0=5k_F$ at the NS
  junctions.  $G_{RL}$ is plotted in the upper panel as a 
  contour-plot in units of $2e^2/h$. In the lower panel, ET and cAR 
  contributions to $G_{RL}$  are plotted at a fixed bias $eV=0\De$ as a
  function of the length of the SC region in units of $k_F^{-1}$. cAR 
  enhancement is $\pi$ periodic in $k_F a$.
  }\label{fig-cAR-NSN}  \end{figure}

\noi Chtchelkatchev~\cite{cht03} has performed calculations for a two-dimensional 
NM-SC-NM system with barriers at the junctions. We study the 
one-dimensional version of the NM-SC-NM set-up with barriers here and
reproduce their results qualitatively, i.e., show that cAR is enhanced by having 
barriers at the NM-SC interface. Fig.~\ref{fig-cAR-NSN} summarizes the
results of our calculations on a single-channel SC connected to two NM
leads with barriers of strength $q_0=5k_F$ (where 
$k_F=\sqrt{2m\mu/\hbar^2}$). The parameter $q_0$ enters the calculations
through the BC: 
\beq \Do_x\psi(x)|_{x=\pm a+0} - \Do_x\psi(x)|_{x=\pm a-0}=
q_{0}~\psi(\pm a).  \eeq
In certain narrow regions in the bias-$k_F a$ 
plane, cAR is enhanced. But the enhanced cAR contributions to the 
conductance are much smaller even in the theoretical calculations
for a ballistic system ($0.15 \cdot 2 e^2/h$ being the largest 
value obtained in this setup as opposed to the maximum possible value 
of $2 e^2/h$). The mechanism here is multiple back-and-forth
reflections of the BdG quasiparticles in the SC due to the 
presence of barriers at the NS junctions. Though the spectrum is 
gapped, the evanescent BdG modes have some real component to their 
``momenta" and that is what amounts to saying {\sl `multiple 
back-and-forth reflections'}.

Experiments so far have found a very small enhancement factor 
for cAR. Russo~et~al~\cite{russo05} have performed experiments 
on a set-up that is qualitatively an NM-SC-NM junction with barriers.
But, this differs from the one calculations have been performed on 
mainly in the fact that the NM's are diffusive while the the calculations have 
been for NM-SC-NM. Thus, the results of the calculations cannot be 
directly used to understand experiments. Nevertheless, evidence of the 
enhancement of cAR is found in experiments and the negative non-local voltage measured is of the 
order  of $10^{-2}$ times the value of the normal state voltage. This is 
to be contrasted  with the value of $0.8$ that we get in our calculations.
The BdG modes also have an imaginary part to the momentum in addition
to the real part which suppresses the non-local subgap transport when
the length of the SC is very large. In the opposite limit of a short 
SC region, the electron to hole conversion is very small. Also, in the 
limit of the barrier strength $q_0$ much smaller than the Fermi 
wavenumber $k_F$, the effect of the barrier becomes negligible and the 
dominance of ET is restored. Ideally, to enhance cAR in this setup, 
$k_F a \sim \pi$ and $q_0 \gtrsim k_F$ to enable  
a few back-and-forth reflections of the BdG quasiparticle modes 
before they decay (the real part of $k$ is responsible for this 
interference) and electron to hole conversion to happen. 

\section{Josephson junction between two-dimensional
superconductors} 
\noi A system closely related to the ladder proposed here is a Josephson
junction between a pair of two-dimensional superconductors. The Hamiltonian
that describes such a Josephson junction is: 
\bea  
H  &=& \sum_{s}\Bigg[\Psi^{\dg}_{s}(x,y) 
\Big\{ -\Big(\f{\hbar^2}{2m} \f{\Do^2}{\Do x^2} + \f{\hbar^2}{2m} 
\f{\Do^2}{\Do y^2} + \mu\Big)\tau_z \nn \\ && 
+ \Delta~\big[\cos{[\phi(y)]}~\tau_x
+ \sin{[\phi(y)]} ~\tau_y\big] \Big\}\Psi_{s}(x,y) \Bigg], \nn \\
 && {\rm where~~}  \phi(y) = 0 ~~{\rm for~~} y>0, \nn \\ &&
 {~\rm and~~~~ }  \phi(y) =  -\phi ~~{\rm for~~} y<0. 
\eea
The dispersion for the bulk as given by this Hamiltonian is
\beq E(\vec k)=\pm\sqrt{(\hbar^2\vec k ^2/2m-\mu)^2+\De^2}, \label{eq-disp2} \eeq 
and the bulk gap is $2\De$.
We show here that  a non-zero $\phi$ can induce one-dimensional states localized
at the junction. The BC for the junction is- $\psi(x,y=0^+)=\psi(x,y=0^-)$ and 
$[\Do_y\psi(x,y)|_{y=0^+}-\Do_y\psi(x,y)|_{y=0^-}]=q_0\psi(x,y=0)$. The parameter 
$q_0$ characterizes the transparency of the junction. The limits $q_0=0$ and 
$q_0\to\infty$ correspond to fully transparent and fully opaque junctions
respectively. The subgap states require at least one out of $k_x$ and $k_y$
to be complex. Since the junction is in the $y$-direction, $k_y$ is complex.
Translational invariance of the system along the $x$-direction makes $k_x$ a good 
quantum number and real valued. At a given energy $E$ in the gap (i.e., 
$|E|<\De$) and in a particular spin eigensector~$s$, the wavefunction is
\bea  \psi(x,y) = e^{i k_x x} \cdot \sum_{\nu=\pm} A_{s_y,\nu} e^{i\nu k_R y 
- s_y \kappa y}~ \vec u_{s_y,\nu}, \eea
where $k_y=\nu k_R + i s_y \kappa$, ($k_R,~\kappa>0$), 
$\vec k = (k_x,k_y)$, $\vec k$ is related to $E$ by the dispersion
relation eq.~\eqref{eq-disp2}, $s_y=sign(y)$, and $\vec u_{s_y,\nu}$
is the eigenspinor that is a function of $k_y=\nu k_R + i s_y \kappa$.
Substitution of the above wavefunction in the BC equation yields
a subgap state that exists if and only if ${\rm Det}[M]=0$, where $M$
is $4\times 4$ matrix, given by:
\bea
M &= &\begin{bmatrix}
       M_1 & M_2
      \end{bmatrix}, {\rm ~ where~} \nn \\
&& M_1 = \begin{bmatrix}
\vec u_{+,+} & \vec u_{+,-} \\
(ik_{+,+}-q_0) \vec u_{+,+} & (ik_{+,-} -q_0) \vec u_{+,-} \end{bmatrix}
\nn \\ && M_2 = \begin{bmatrix} \vec u_{-,+} & \vec u_{-,-} \\
 -ik_{-,+}~ \vec u_{-,+} & -ik_{-,-}~ \vec u_{-,-} 
\end{bmatrix}. \label{eq-M} 
\eea
\begin{figure}[htb]
\includegraphics[width=8cm]{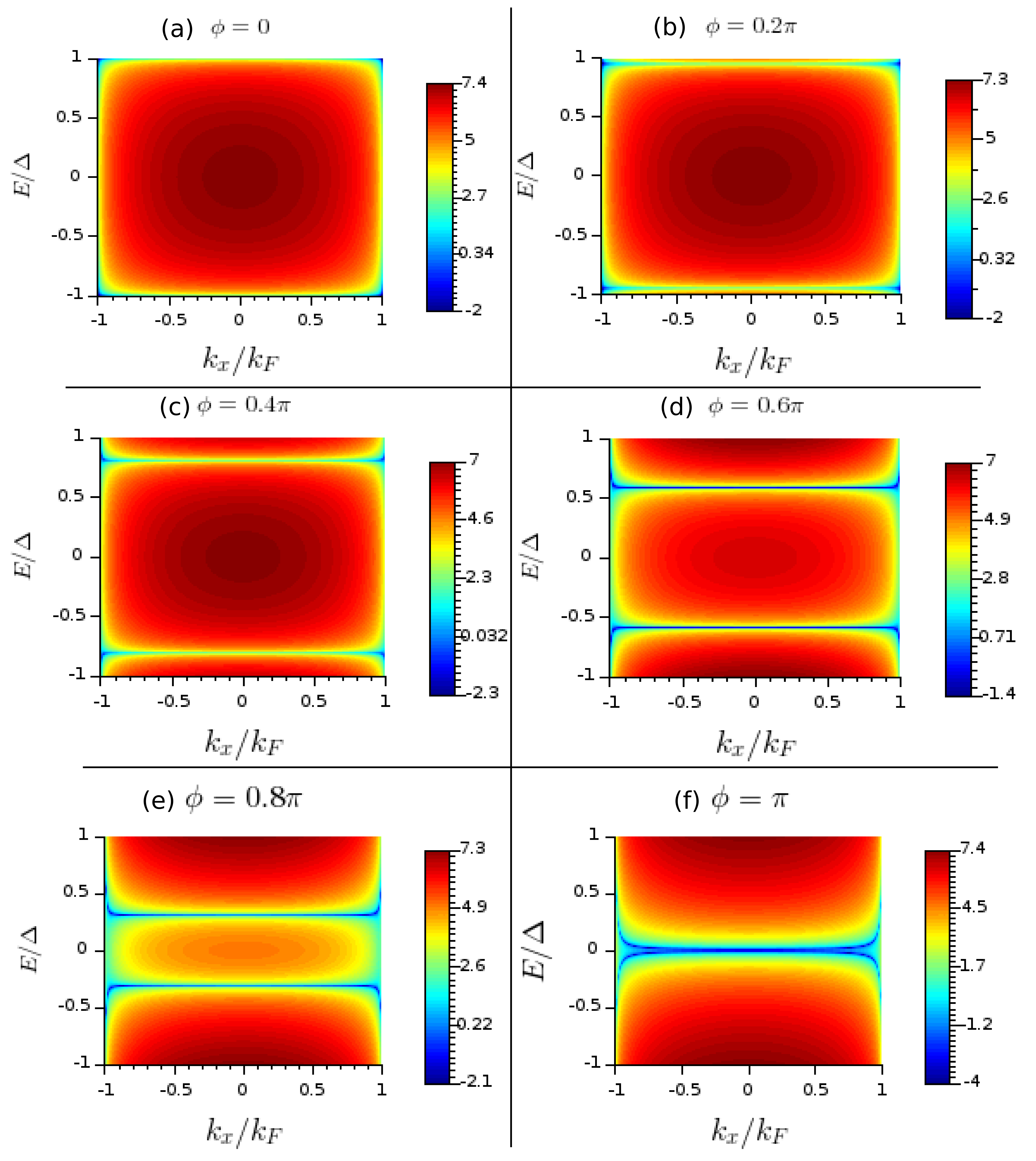}
\caption{$\log|{\rm Det.}[M]|$ plotted as a function of $E$ and $k_x$ 
for various choices of $\phi$. Dark blue regions in each contour plot
indicate the existence of subgap states localized along the junction.
We have chosen $q_0=0$ everywhere.}
 ~\label{fig-2djj} \end{figure}
The results of a numerical calculation presented in Fig.~\ref{fig-2djj} show that a nonzero 
phase difference $\phi$ results in subgap states, that go deeper into the gap 
as $\phi$ approaches $\pi$. The dispersion of the 1D modes is almost flat in
 the middle, while for $k_x$ near $\pm k_F$, the dispersion has a sharp slope and 
 connects to energies $\pm \Delta$ smoothly.

\end{document}